\documentclass[prb,aps,twocolumn,showpacs,preprintnumbers,amsmath,amssymb,superscriptaddress,floatfix]{revtex4}

\usepackage{graphicx}
\usepackage{dcolumn}
\usepackage{bm}
\usepackage{natbib}
\usepackage{amsmath}
\usepackage{color}

\begin{document}

\title{Spatially--resolved analysis of edge--channel equilibration in quantum Hall circuits}

\author{Nicola Paradiso}
\affiliation{NEST, Istituto Nanoscienze--CNR and Scuola Normale Superiore, Pisa (PI), Italy}

\author{Stefan Heun}\email{stefan.heun@nano.cnr.it}
\affiliation{NEST, Istituto Nanoscienze--CNR and Scuola Normale Superiore, Pisa (PI), Italy}

\author{Stefano Roddaro}
\affiliation{NEST, Istituto Nanoscienze--CNR and Scuola Normale Superiore, Pisa (PI), Italy}

\author{Davide Venturelli}
\affiliation{NEST, Istituto Nanoscienze--CNR and Scuola Normale Superiore, Pisa (PI), Italy}
\affiliation{Institut NEEL, CNRS and Universit\'{e}  Joseph Fourier, Grenoble, France}

\author{Fabio Taddei}
\affiliation{NEST, Istituto Nanoscienze--CNR and Scuola Normale Superiore, Pisa (PI), Italy}

\author{Vittorio Giovannetti}
\affiliation{NEST, Istituto Nanoscienze--CNR and Scuola Normale Superiore, Pisa (PI), Italy}

\author{Rosario Fazio}
\affiliation{NEST, Istituto Nanoscienze--CNR and Scuola Normale Superiore, Pisa (PI), Italy}

\author{Giorgio Biasiol}
\affiliation{Istituto Officina dei Materiali CNR, Laboratorio TASC, Basovizza (TS), Italy}

\author{Lucia Sorba}
\affiliation{NEST, Istituto Nanoscienze--CNR and Scuola Normale Superiore, Pisa (PI), Italy}

\author{Fabio Beltram}
\affiliation{NEST, Istituto Nanoscienze--CNR and Scuola Normale Superiore, Pisa (PI), Italy}

\date{\today}

\begin{abstract}
We demonstrate an innovative quantum Hall circuit with variable geometry employing the moveable electrostatic potential induced by a biased atomic force microscope tip. We exploit this additional degree of freedom to identify the microscopic mechanisms that allow two co--propagating edge channels to equilibrate their charge imbalance. Experimental results are compared with tight--binding simulations based on a realistic model for the disorder potential. This work provides also an experimental realization of a beam mixer between co--propagating edge channels, a still elusive building block of a recently proposed new class of quantum interferometers.
\end{abstract}


\maketitle


\section{Introduction}

Suppression of backscattering and a very large coherence length are the characteristic properties of edge states~\cite{Halperin1982} in the quantum Hall (QH) regime at the basis of the newly--developed quantum electron interferometry. In this field a number of breakthroughs have appeared in recent years, such as the experimental realization of 
Mach--Zehnder,\cite{Ji2003,Neder2006,Neder2007,Roulleau2007} Fabry--P\'erot,\cite{Camino2005} and  Hanbury--Brown--Twiss~\cite{NederNature2007} electron interferometers.   In these devices the electronic analogue of a beam splitter is obtained by a quantum point contact, a powerful tool which we have recently used  to study the electron tunneling between \textit{counter--propagating} edge states.\cite{Roddaro2003,Roddaro2004,Roddaro2005,Roddaro2009}
The constantly growing flexibility in the practical realization of QH nanostructures stimulates further investigations and  different designs that are often inspired by quantum optics. One particularly appealing possibility is to exploit interference of \textit{co--propagating} edge
channels since it allows the concatenation of several 
interferometers.\cite{Giovannetti2008} Within this architecture, a beam--splitter can be realized by sharp, localized potentials capable of inducing coherent inter--channel scattering, see e.g.~Refs.~\onlinecite{Tejedor1991,Palacios1992,Palacios1993,ventu,Olendski}. Appropriate design of such interferometers requires the detailed understanding of the physics of co--propagating edges. 

Several groups~\cite{Muller1992,Komiyama1992,Acremann1999,Wurtz2002,Nakajima2010} measured charge transfer and the electro--chemical potential imbalance equilibration between co--propagating edge channels. M\"uller \textit{et al.}~\cite{Muller1992} and W\"urtz \textit{et al.}~\cite{Wurtz2002} interpreted their results in terms of classical rate equations, while only very recently  the contribution of coherent effects in the equilibration process has been considered.\cite{Deviatov2008,Deviatov2009} In these experiments, two co--propagating edge channels originating from two ohmic contacts at different potential meet at the beginning of a common path of fixed length $d$ where charge transfer tends to equilibrate their voltage difference.\cite{Wurtz2002} At the end of the path the edge channels are separated by a selector gate and guided to two distinct detector contacts.  Consequently, while these setups  yield valuable information on the {\em cumulative} effect of the processes taking place along the whole distance $d$, they make it impossible to link charge transfer to local sample characteristics. 

\begin{figure}[tbp]
\includegraphics[width=\columnwidth]{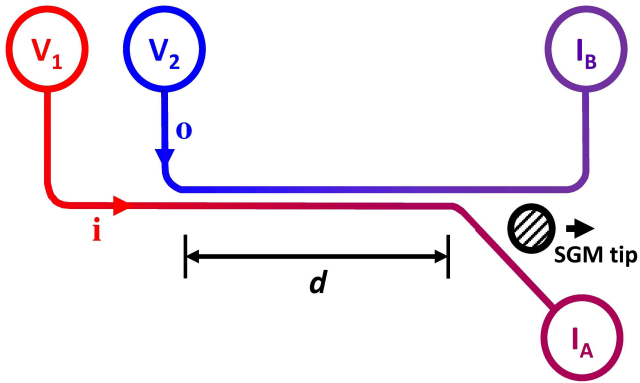}
\caption{(Color online) Schematic drawing of the key idea behind our experiment: the SGM tip is used to actively control the edge trajectories so to obtain a continuously tunable interaction region length $d$. This allows a spatially--resolved analysis of the equilibration process.}
\label{fig:sketch}
\end{figure}

In order to shed light on this issue, 
in this article we present a novel approach to Scanning Gate Microscopy (SGM) that allows us to 
investigate  the spatial evolution of the inter--channel scattering between co--propagating edge states 
in the QH regime with unprecedented spatial resolution.  Here, the SGM tip is used not merely as a probe, but as an active component of a complex device which permits one to address quantum structures whose dimensions can be  tuned during the measurement. For this purpose, we implemented a special QH circuit with variable geometry, in which the length of the interaction path can be continuously changed by positioning the biased tip of the SGM  (see Fig.~\ref{fig:sketch}). This  movable tip  introduces a new degree of freedom for transport experiments, since it allows to continuously control the size of a single component of the device under investigation during the same low--temperature measurement session. 
For large values of $d$ our findings are consistent with the results of Refs.~\onlinecite{Muller1992,Komiyama1992,Wurtz2002,Nakajima2010},  i.e. the bias imbalance shows an exponential decay whose characteristic length is the  equilibration length $\ell_{eq}$. For small $d$, however, we are able to reveal by a direct imaging technique the effect of individual scattering centers in tranferring electrons among co--propagating edges. Numerical simulations of the device based on the
Landauer--B\"uttiker formalism~\cite{buttiker,buttikerbis}  show that inter--channel scattering can occur while coherence is maintained,
suggesting  the possibility  that such  mechanisms could be used 
as the basic ingredient to  build simply--connected, easily--scalable interferometers along the lines proposed in Ref.~\onlinecite{Giovannetti2008}.

\section{Experimental Details}

The samples for this study were fabricated starting from an Al$_{0.33}$Ga$_{0.67}$As/GaAs heterostructure with a two--dimensional electron gas (2DEG) which is confined 55~nm underneath the surface. Its electron sheet density and mobility at low temperature are $n=3.2\cdot 10^{15}$~m$^{-2}$ and $\mu=4.2 \cdot 10^2$~m$^{-2}$/Vs respectively, as determined by Shubnikov--de Haas measurements.

\begin{figure}[tbp]
\includegraphics[width=\columnwidth]{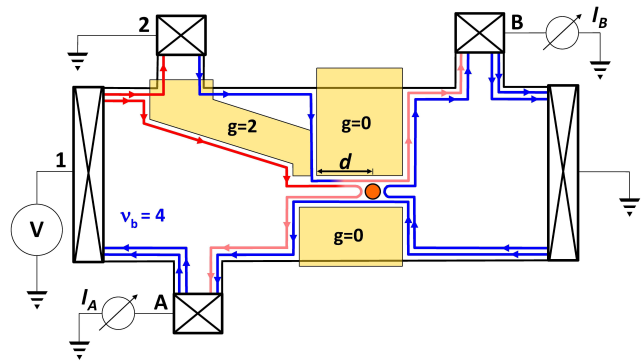}
\caption{(Color online) Scheme of the experimental setup.  Three Schottky gates are used to independently contact two co--propagating edge channels and to define a 6 $\mu$m long and 1 $\mu$m wide constriction. Using the SGM tip it is possible to selectively reflect the inner channel and define a variable inter--edge relaxation region length $d$.}
\label{fig:setup}
\end{figure}

The Hall bar was patterned via optical lithography and wet etching. Ohmic contacts were obtained by evaporation and thermal annealing of a standard Ni/AuGe/Ni/Au multilayer (10/200/10/100~nm). All gates were defined by electron beam lithography and consist of a Ti/Au bilayer (10/20~nm). Two nominally identical devices (S1 and S2) were produced as outlined in Fig.~\ref{fig:setup}.

Our measurements were performed with the 2DEG at bulk filling factor $\nu_b=4$  ($B=3.32$ T). At such field, the effective distance between 
edge states separated by the cyclotron gap ($\hbar \omega_c=5.7$~meV)  is of the order of 100~nm, as we showed in our previous measurements on a similar sample.\cite{Paradiso2010} In general, in a sample with a given confinement profile the inter--edge channel distance is proportional to the energy gap between Landau levels. Since the Zeeman gap is of the order of 0.1~meV (we assume $g^{\ast}=-0.44$),\cite{Thomas1998} the distance between Zeeman--split edge states is so small that they cannot be resolved in our experiment. Thus here we consider pairs of Zeeman--split edges as one individual channel carrying $2G_0\equiv2e^2/h$ units of conductance. Finally, since we work at $\nu_b=4$, two spin--degenerate edge channels are populated.

The SGM system is mounted on the cold finger (base temperature 300 mK) of a $^3$He cryostat.\cite{Paradiso2010} The sample temperature, calibrated with a Coulomb blockade thermometer, is 400 mK. The maximum scanning area of the SGM at 300 mK is 8.5~$\mu$m $\times$ 8.5 $\mu$m. The coarse and fine control of the tip--sample position is provided by a stack of piezo--actuators. The sample is mounted on a chip carrier positioned on top of the piezo--scanner. The SGM tip was obtained by controlled etching of a  $50$~$\mu$m thick tungsten wire. This resulted in tips with a typical radius of about 30 nm. The tip was then glued on a quartz tuning fork, which allowed us to perform topography scans by controlling the oscillation amplitude damping due to the tip--sample shear force. Due to the close tip--sample proximity, during the topography scans both the tip and the gates are temporarily grounded in order to avoid shorts. On the other hand, during the SGM measurements the tip  (biased at the voltage $V_{tip}=-10 V$) is scanned  about 40~nm above the heterostructure surface, in order to  avoid both accidental contacts between the biased tip and the gates and to keep the tip--2DEG distance constant, irrespective of the topographic details.

The cryostat is equipped with a superconducting magnet coil which provides magnetic fields up to 9~T. The whole setup is decoupled from the lab floor by means of a system of springs in order to damp mechanical noise. Images are processed with the WSxM software.\cite{Horcas2007} In all  conductance maps shown in this article, the effect of the series resistance of both the external wires and the ohmic contacts has been subtracted.

The geometry of the QH circuit is determined by the electrostatic potential induced by three Schottky gates and the SGM tip. The upper left gate in Fig.~\ref{fig:setup} defines a region with local filling factor $g=2$ which selects only one of the two channels propagating from contact 1 at voltage $V$ and guides it towards 
contact 2. When this is grounded, an imbalance is established between edge channels at the entrance of the constriction defined by the two central gates at local filling factor $g=0$. The two channels propagate in close proximity along the constriction, which is 6~$\mu$m long and 1~$\mu$m wide. In our experiments, we suitably positioned the depletion spot induced by the biased tip of the SGM so that the inner channel is completely backscattered, while the outer one is fully transmitted. As a consequence, the two channels are separated after a distance $d$ that can be adjusted by moving the tip. Since the outer edge was grounded before entering the constriction, the detector contact B will measure only the electrons scattered between channels, while the remaining current is detected at contact A. 

\section{Results}

\begin{figure}[tbp]
\centering
\includegraphics[height=0.76\textheight]{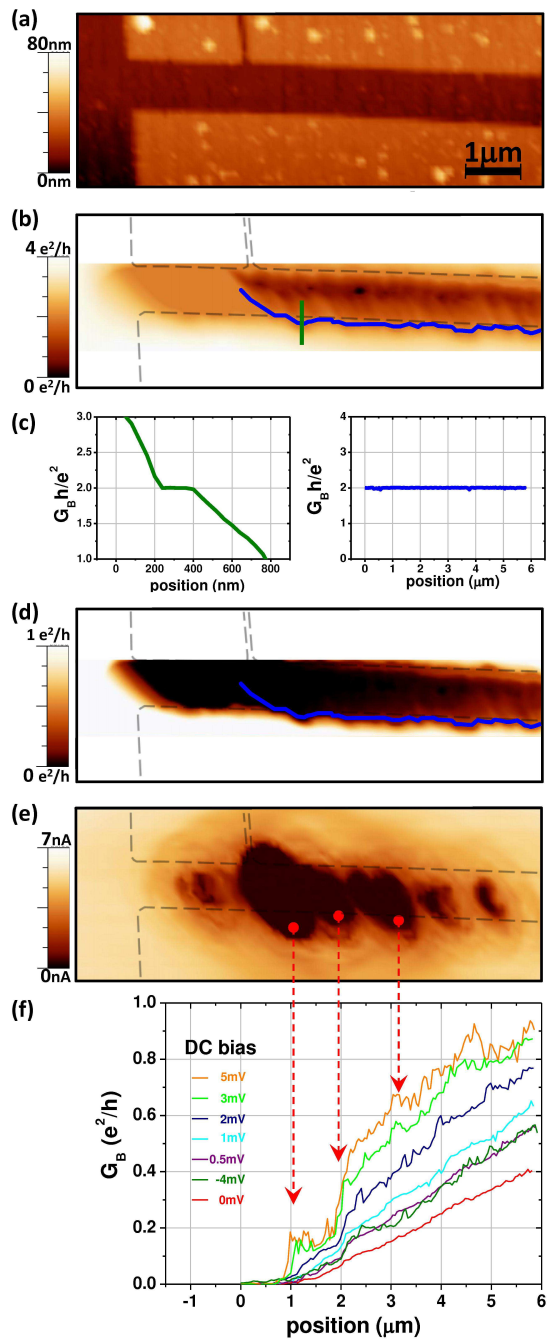}
\caption{(Color online) (a) Topography scan of  device S1. (b) Calibration scan: the SGM map refers to the differential conductance signal measured at  contact B when contact 2 is floating. $V_{tip}=-10$ V. (c) Conductance profiles measured along the green (left panel) and the blue (right panel) line in (b). (d) Imaging of the inter--channel equilibration (contact 2 grounded). (e) SGM measurement at zero magnetic field, with DC source bias $V=100$ $\mu$V. (f) Finite bias equilibration signal measured along the trajectory determined by means of the calibration scan. There is a clear correlation between the steps in the equilibration curves and the position of scattering centers in the SGM scan at zero magnetic field.  Furthermore, we observe an enhancement of the equilibration steps with increasing bias.} 
\label{fig:stack}
\end{figure}

The peculiar geometry of this QH circuit implies that all measurements critically depend on the ability to set the edge configuration so that the inner edge is perfectly reflected while the outer one is fully transmitted. To this end, we first performed topography scans (Fig.~\ref{fig:stack}a) yielding a reference frame to evaluate the relative position of the tip with respect to the confining gates in the subsequent SGM scans.
Then we performed calibration scans aimed at establishing tip trajectories ensuring that the inner channel is indeed completely backscattered, while the outer one is fully transmitted (edge configuration as sketched in Fig.~\ref{fig:setup}). In these scans, a small AC bias (50~$\mu$V) was applied to  source contact 1, while contact 2 was kept floating so that both channels at the entrance of the central constriction are at the same potential and carry the same current $I_1=I_2=2G_0V$. 
Fig.~\ref{fig:stack}b shows a map of the differential conductance $G_B=\partial I_B/\partial V$ measured at contact~B by standard lock--in technique and obtained by scanning the biased tip inside the constriction. The color plot of Fig.~\ref{fig:stack}b can be interpreted as follows: when the tip is far from the constriction axis both channels are fully transmitted to the drain contact B and the measured total conductance is $G_B=4G_0$. By moving the tip towards the axis of the 1D--channel, the inner edge channel is increasingly backscattered and the conductance decreases until we reach a plateau for $G_B=2G_0$ (left panel of Fig.~\ref{fig:stack}c). This plateau is due to the spatial separation $\delta$ between the two edge channels. In fact once the inner channel is completely backscattered, it is necessary to move the tip approximately $2\delta$ further before reflection of the outer one occurs, as discussed in Refs.~\onlinecite{Aoki2005,Paradiso2010}. Thus the tip trajectory ensuring the desired edge configuration (Fig.~\ref{fig:setup}) was determined as the locus of the middle points of the plateau strip (blue line in Fig.~\ref{fig:stack}b). As shown in the right panel of Fig.~\ref{fig:stack}c, the conductance along this trajectory is constant and equals the conductance of a single channel, i.e.~$2G_0$.

Next, we imaged the inter--channel differential conductance. The two edge channels entering the constriction were imbalanced by grounding contact 2. In this configuration, at the beginning of the interaction path, only the inner channel carries a non--zero current, i.e.~$I_1=2G_0V$, where $V$ is the source voltage. The electrochemical potential balance is gradually restored by scattering events that take place along the interaction path, which yields a partial transfer of the initial current signal from the inner to the outer channel. The device architecture allowed us to detect both transferred electrons and reflected ones by measuring the current signal at contacts B and A, respectively. 
We verified that the sum of currents measured at A and B is constant and always equal to $2G_0V$.

Figure~\ref{fig:stack}d shows the SGM map of the inter--channel differential conductance $G_B$ at zero DC bias. The key feature of this scan is the monotonic increase of the scattered current as a function of the interaction distance $d$.
This can be directly observed in Fig.~\ref{fig:stack}f, where we show several finite--bias conductance profiles acquired along the trajectory determined in the previous calibration step. For a given value of $d$, the dramatic enhancement of the equilibration for finite DC bias is consistent with the results obtained by means of I--V characteristics in samples with fixed interaction length.\cite{Wurtz2002} In particular, for DC bias of the order of the cyclotron gap, $\hbar\omega_c=5.7$ meV, the differential conductance reaches its saturation value $G_B=G_0$, which corresponds to a transmission probability $T_{12}=0.5$, i.e.~$I_A = I_B$.

All curves in Fig.~\ref{fig:stack}f are characterized by sharp steps in some positions. This behavior was confirmed by measurements on other devices, which showed the same stepwise monotonic behavior albeit with different step positions. This indicates that the scattering probability is critically influenced by local details of each sample, e.g.~by the location of impurities that can produce sharp potential profiles whose effect in the QH inter--channel scattering can be revealed by the SGM technique.\cite{Woodside2001}
In order to correlate the presence of scattering centers with the steps in the conductance profile we performed SGM scans at zero magnetic field (Fig.~\ref{fig:stack}e). Such a scan provides a direct imaging of the disorder potential and can identify the most relevant scattering centers (see Refs.~\onlinecite{Topinka2001,Steele2005} for similar scanning probe microscopy investigations).
A comparison between Fig.~\ref{fig:stack}e and Fig.~\ref{fig:stack}f shows a clear correlation between the steps in the conductance profiles with the main spots in the disorder--potential map. 
This is the central finding of the present work and establishes a direct link between the atomistic details of the sample and the inter--channel transport characteristics. Such correlation is impossible to detect with standard transport measurements and requires the use of scanning probe microscopy techniques.
  
It is important to note that inter--channel transmission is nearly zero up to the first scattering center. This indicates that impurity--induced scattering is the dominant process equilibrating the imbalance, while other mechanisms that were invoked in literature, such as the acoustic--phonon scattering, have only a negligible effect for short distances,  in agreement with the theoretical findings of Ref.~\onlinecite{Komiyama1992}. We also observe that the step amplitude is suppressed when the length of the interaction path $d$ is bigger than about 3~$\mu$m.

\section{Discussion}

In view of possible applications to QH interferometry, it is necessary to determine the degree of coherence of the position--dependent, inter--channel differential conductance.
For this reason we make use of a theoretical model which accounts for elastic scattering only and restrict our analysis to the zero--DC bias case.
The system is described through a tight--binding Hamiltonian, where the magnetic field is introduced through Peierls phase factors in the hopping potentials.
According to the Landauer--B\"uttiker formalism,\cite{buttiker,buttikerbis} the differential conductance is determined by the scattering coefficients which are calculated using a recursive Green's function technique.
Apart from a hard--wall confining potential, electrons are subjected to a disorder potential consisting of few strong scattering centers on top of a background potential.
Scattering centers are modeled as Gaussian potentials whose positions (which are different from device to device) are deduced from SGM scans in the constriction at zero magnetic field (Fig.~\ref{fig:stack}e shows one example). The height of the Gaussian potentials is of the order of the cyclotron gap and their spatial variation occurs on a length scale of the order of the magnetic length ($\ell_B \approx$ 15~nm).
The background potential is modeled as a large number of randomly distributed smooth Gaussian potentials, whose height is of the order of one tenth of the cyclotron gap. The conductance is finally calculated averaging over a large number of random configurations of the background potential to account for phase–-averaging mechanisms which are always present in the system.

\begin{figure}[tbp]
\centering
\includegraphics[width=\columnwidth]{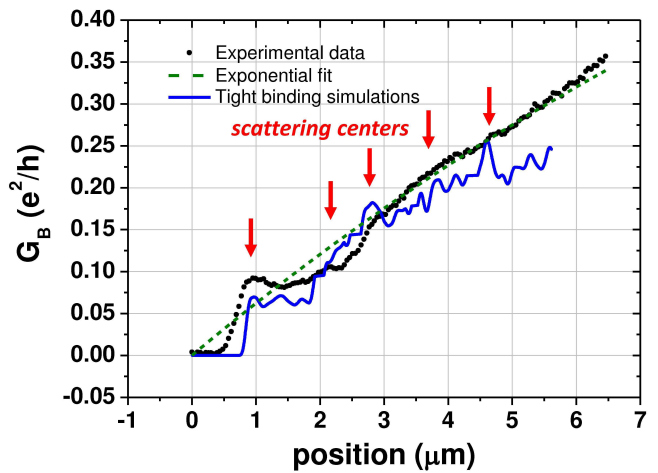}
\caption{(Color online) Results of the tight--binding simulations  for the zero--bias case: the inter--channel, zero--temperature differential conductance (solid line) compared with experimental data  from device S2 (filled dots). From  the exponential fit (dashed line) we deduce an equilibration length $\ell_{eq}=15$~$\mu$m. The position of strong scattering centers in the simulation is indicated by red arrows.  Comparison of the curves in Figs.~\ref{fig:stack} and \ref{fig:tb} demonstrates that the position of the jumps changes from sample to sample and critically depends on the specific distribution of the scattering centers in each sample, which is the main finding of our article.}
\label{fig:tb}
\end{figure}

Figure~\ref{fig:tb} shows results of our simulations (solid blue line), together with the experimental data from device S2 for $V=0$ (filled black dots) and an exponential fit (dashed green line). For short distances the computed conductance exhibits steps in correspondence to the scattering centers (positions indicated by red arrows in Fig.~\ref{fig:tb}), while 
at larger distances it presents a monotonic behavior where the steps are washed out by the averaging over the background. Both regimes are consistent with the experimental data.
 
In Fig.~\ref{fig:tb} we also compare our experimental data with the exponential behavior $G_B=G_0(1-e^{-d/\ell_{eq}})$ which was reported previously.\cite{Muller1992,Wurtz2002} For short $d$, there is a discrepancy between the experimental conductance profile and the exponential curve, due to the discreteness of the scattering centers. On the other hand, for larger distances our experimental data are well fitted by the exponential curve. 
We would like to underline that here we actually directly verify this exponential behavior, by continuously tuning the interaction length~$d$. In previous works, the equilibration length $\ell_{eq}$ was extracted from four-wire resistance measurements at fixed~$d$, \textit{assuming} an exponential dependence.\cite{Muller1992,Komiyama1992,Wurtz2002,Machida1996} From our data we obtain an equilibration length $\ell_{eq}=15$~$\mu$m, which is of the same order of magnitude as values reported in literature.\cite{Machida1996} 

We also performed  measurements at bulk  filling factor  $\nu_b = 2$, so that the electron transfer takes place between two spin-split edge channels. In this case we did not observe equilibration at zero-bias, consistent with the  fact that  typical equilibration lengths reported in literature for  $\nu_b = 2$ are of the order of millimeters.\cite{Muller1992} In view of possible applications as beam splitter it is therefore advantageous to work at  $\nu_b = 4$  since one needs to achieve a coherent mixing with an interaction  path as short as possible.

In conclusion, we used the biased tip of a SGM as an active component of a QH circuit which implements a tunable beam splitter to mix co--propagating edge states. The ability to control the interaction path length allowed us to identify the microscopic mechanisms governing inter--channel electron scattering. From the comparison of several conductance profiles (as the one shown in  Fig.~\ref{fig:stack}f) acquired with different devices, we can conclude that scattering induced by impurities is the key process that enables charge transfer between the channels. This conclusion is supported by theoretical simulations. This allows application of this device as a beam splitter in the simply--connected Mach--Zehnder interferometer proposed in Ref.~\onlinecite{Giovannetti2008} and opens new possibilities in quantum electron interferometry.

\begin{acknowledgments}
We acknowledge financial support from the Italian Ministry of
Research (FIRB projects RBIN045MNB and RBID08B3FM).
\end{acknowledgments}


\begin{thebibliography}{99}

\bibitem{Halperin1982} B.~I. Halperin, Phys. Rev. B {\bf 25}, 2185 (1982).

\bibitem{Ji2003} Y. Ji, Y. Chung, D. Sprinzak, M. Heiblum, D. Mahalu, and H. Shtrikman, Nature {\bf 422}, 415 (2003). 

\bibitem{Neder2006} I. Neder, M. Heiblum, Y. Levinson, D. Mahalu, and V. Umansky, Phys. Rev. Lett. {\bf 96}, 016804 (2006).

\bibitem{Neder2007} I. Neder, F. Marquardt, M. Heiblum, D. Mahalu, and V. Umansky, Nat. Phys. {\bf 3}, 534 (2007).

\bibitem{Roulleau2007} P. Roulleau, F. Portier, D. C. Glattli, P. Roche, A. Cavanna, G. Faini, U. Gennser, and D. Mailly, Phys. Rev. B {\bf 76}, 161309(R) (2007).

\bibitem{Camino2005} F.~E. Camino, W. Zhou, and V.~J. Goldman, Phys. Rev. Lett. {\bf 95}, 246802 (2005).

\bibitem{NederNature2007} I. Neder, N. Ofek, Y. Chung, M. Heiblum, D. Mahalu, and V. Umansky, Nature {\bf 448}, 333 (2007). 

\bibitem{Roddaro2003} S. Roddaro, V. Pellegrini, F. Beltram, G. Biasiol, L. Sorba, R. Raimondi, and G. Vignale, Phys. Rev. Lett. {\bf 90}, 046805 (2003).  

\bibitem{Roddaro2004} S. Roddaro, V. Pellegrini, F. Beltram, G. Biasiol, and L. Sorba, Phys. Rev. Lett. {\bf 93}, 046801 (2004).

\bibitem{Roddaro2005} S. Roddaro, V.  Pellegrini, F. Beltram, L.~N. Pfeiffer,  and K.~W. West, Phys. Rev. Lett. {\bf 95}, 156804 (2005). 

\bibitem{Roddaro2009} S. Roddaro, N. Paradiso, V. Pellegrini, G. Biasiol, L. Sorba, and F. Beltram, Phys. Rev. Lett. {\bf 103}, 016802 (2009).  

\bibitem{Giovannetti2008} V. Giovannetti, F.  Taddei, D.  Frustaglia,  and R. Fazio, Phys. Rev. B {\bf 77}, 155320 (2008).

\bibitem{Tejedor1991} C. Tejedor and J.~J. Palacios, Physica Scripta {\bf T35}, 121 (1991).

\bibitem{Palacios1992} J.~J. Palacios  and  C. Tejedor, Phys. Rev. B {\bf 45}, 9059 (1992).

\bibitem{Palacios1993} J.~J. Palacios  and  C. Tejedor, Phys. Rev. B {\bf 48}, 5386 (1993).

\bibitem{Olendski} O. Olendski and L. Mikhailovska, Phys. Rev. B {\bf 72}, 235314 (2005). 

\bibitem{ventu} D. Venturelli, V. Giovannetti, F. Taddei, R. Fazio, D. Feinberg, G. Usaj, and C. A. Balseiro, \textit{preprint arXiv}: 1008.1913.

\bibitem{Muller1992} G. M\"uller, D. Weiss, A. V. Khaetskii, K. von Klitzing, S. Koch, H. Nickel, W. Schlapp, and R. L\"osch, Phys. Rev. B {\bf 45}, 3932 (1992).

\bibitem{Komiyama1992} S. Komiyama, H. Hirai, M. Ohsawa, Y. Matsuda, S. Sasa, and T. Fujii, Phys. Rev. B {\bf 45}, 11085 (1992).

\bibitem{Acremann1999} Y. Acremann, T. Heinzel, K. Ensslin, E. Gini, H. Melchior, and M. Holland, Phys. Rev. B {\bf 59}, 2116 (1999).

\bibitem{Wurtz2002} A. W\"urtz, R. Wildfeuer, A. Lorke, E.~V. Deviatov,  and V.~T. Dolgopolov, Phys. Rev. B {\bf 65}, 075303 (2002).

\bibitem{Nakajima2010} T. Nakajima, Y. Kobayashi, S. Komiyama, M. Tsuboi, and T. Machida, Phys. Rev. B {\bf 81}, 085322 (2010).



\bibitem{Deviatov2008} E. V. Deviatov and A. Lorke, Phys. Rev. B \textbf{77}, 161302(R) (2008).

\bibitem{Deviatov2009} E. V. Deviatov, B. Marquardt, A. Lorke, G. Biasiol, and L. Sorba, Phys. Rev. B \textbf{79}, 125312 (2009).

\bibitem{buttiker}  M. B\"{u}ttiker, Y. Imry, R. Landauer,  and S. Pinhas,  Phys. Rev. B \textbf{31}, 6207 (1985).

\bibitem{buttikerbis} M. B\"{u}ttiker,  Phys. Rev. B \textbf{38}, 9375 (1988).

\bibitem{Paradiso2010} N. Paradiso, S. Heun, S. Roddaro, L. N. Pfeiffer, K. W. West, L. Sorba, G. Biasiol, F. Beltram, Physica E {\bf 42}, 1038 (2010).

\bibitem{Thomas1998} K. J. Thomas, J. T. Nicholls, N. J. Appleyard, M. Y. Simmons, M. Pepper, D. R. Mace, W. R. Tribe, and D. A. Ritchie, Phys. Rev. B \textbf{58}, 4846 (1998).

\bibitem{Horcas2007} I. Horcas, R. Fernandez, J. M. Gomez-Rodriguez, J. Colchero, J. Gomez-Herrero, and A. M. Baro, Rev. Sci. Instrum. {\bf 78}, 013705 (2007). 

\bibitem{Aoki2005} N. Aoki, C. R. da Cunha, R. Akis, D. K. Ferry,  and Y. Ochiai, Phys. Rev. B \textbf{72}, 155327 (2005).

\bibitem{Woodside2001} M. T. Woodside, C. Vale, P. L. McEuen, C. Kadow, K. D. Maranowski, and A. C. Gossard, Phys. Rev. B {\bf 64}, 041310(R) (2001).

\bibitem{Topinka2001} M. A. Topinka, B. J. LeRoy, R. M. Westervelt, S. E. J. Shaw, R. Fleischmann, E. J. Heller, K. D. Maranowskik, and A. C. Gossard, Nature {\bf 410}, 183 (2001).

\bibitem{Steele2005} G.~A. Steele,  R.~C. Ashoori, L.~N. Pfeiffer,  and K.~W. West, Phys. Rev. Lett. {\bf 95}, 136804 (2005).

\bibitem{Machida1996} T. Machida, H. Hirai, S. Komiyama, T. Osada,  and Y. Shiraki, Phys. Rev. B {\bf 54}, 14261 (1996).

\end{thebibliography}
\end{document}